# Understanding the Branly effect.

**Charles Hirlimann**

Institut de Physique et Chimie des Matériaux de Strasbourg (IPCMS)

ULP-CNRS (UMR 7504), 23 rue du Lœss BP 43, F-67034 Strasbourg cedex 2, France

**Abstract**-At the end of the nineteenth century Édouard Branly discovered that the electrical resistance of a granular metallic conductor could drop by several orders of magnitude when excited by the electromagnetic field emitted by an electrical spark [1]. Despite the fact that this effect has been used to detect radio waves in the early times of wireless telegraphy and more recently studied in the field of granular materials, no satisfactory explanation of the physical origin of the effect has been given yet. In this contribution we propose to relate the Branly effect to the induced tunnelling effect first described by François Bardou and Dominique Boosé [2].

The discovery and characterization of the electromagnetic waves by Hertz at the end of the nineteenth century ignited the still actual studies of the interactions of these new waves with materials. One of the first major findings along this track has been the radioconduction of granular conductors by Branly in 1890 [1]. When an insulator tube filled with metal filings is submitted to a potential difference and receives the transient electromagnetic waves generated by an electrical spark at some distance, its electrical resistance drops down by several orders of magnitude. Many aspects of this effect have been revealed along the following decade. It has been firmly established that the presence of a thin resistive layer between the grains is necessary to achieve a high resistance before the exposure to the electromagnetic waves. As the effect does not occur with noble metal grains that have been cleaned from any surface contaminant, the nature of the resistive layer has been recognized to be an oxide or vacuum by Dorn [3]. It has been early observed that a permanent and unique percolation path is the reason of the drop in resistance of the granular medium [4,5] leading to the understanding that welding of the grains was the result of the coherer effect. Because of this welding of the grains the effect is not reversible although the electrical path can be easily broken by gently striking the tube containing the fillings, restoring the original large resistance of the granular medium. This effect was used by Lodge [6,7] to devise a practical radio detector that insured the pioneering developments of radiotelegraphy. Pressing the grains one against the other reduces the original resistance, increasing the sensitivity of the radioconductor to the external excitation [8]. It was always observed that the effect is inherently noisy, the final resistance fluctuating among a rather large range of results [9]. Above some critical applied potential, the resistance of a granular medium drops down without any external excitation [3]. This effect, must not been confused with the Branly effect and for that reason has been, in the recent times, called the continuous Branly effect [10]. Ashkinass, studying the resistance drop of a unique contact between two tin needles, found a critical voltage of 0.2 V independent of the courant [11]. Guthe and Towbridge recorded the U(I) curve of the contact between two metallic spheres and showed that, without any external excitation, it exhibits a critical voltage P that characterizes the resistance transition of the metal-insulator-metal (MIM) system. They found the value P = 0.23 V for the potential drop between two centimetres sized steel balls.

Interest for the effect then resumed during the last decades in relation with a renewed interest for granular media. The uniqueness of the conduction path was confirmed using infrared optical imaging [12]. The welding of the grains has been firmly and quantitatively established. Careful experiments, in the CW mode, on the conduction between metal spheres has shown, through sophisticated modelling, that high electric currant through contacts with a 100 nm diameter produces a temperature as high as 1050° C, large enough to ensure welding between the metallic conductors [10]. The Branly effect itself was observed in ordered lattices of lead beads by Dorbolo et al [13].

Nevertheless, despite the great wealth of results that has been collected for more than a century, no physical understanding has been proposed yet on the origin of the effect. It is the aim of this paper to give an explanation of the Branly effect in terms of the induced tunnelling effect that has been recently theoretically described and observed [2,14].

The discussion can be first simplified by noting that the grained structure of the conducting medium and the related percolation effects are not relevant to the radio-conduction effect as Branly himself showed in a very elegant experiment performed with a couple of conductors separated by a thin oxide layer [15]. We can assert then that the Branly effect is governed by the electrical tunnel effect that occurs in what is nowadays called a MIM structure. In a typical experiment, Branly observed that when, in the surroundings, a spark emits electromagnetic waves, the electrical resistance of a granular medium drops down. In these experiments a Daniell battery applies a U = 1 V potential difference to the medium and a galvanometer shows the current increase. An extreme drop of resistance was measured to be $R_0$ = 106 Ω to R = 100 Ω. According to Omh's law the original current flowing through the granular medium in that case is $i_0$=10-6 C.s$^{-1}$ corresponding to a flow of electrons equal to $n_0$= 6.10$^{13}$ s$^{-1}$. This current increases up to $i_f$ =10$^{-2}$ C.s$^{-1}$ = 10 mA when the Branly effect takes place. Recently Falcon et al carried out very careful experiments with stainless steel balls in the CW regime and found a very similar value P = 0.4 V per contact for the critical voltage a value in very good agreement with the earlier observation. These authors were able to establish through modelling that the drop of resistance is a consequence of a local heating with a peak temperature of 1050°C, high enough to allow for the local destruction of the barrier of potential and the local welding of the steel balls. Therefore when the current flowing through the MIM contacts of a coherer is forced to increase, a threshold is reached such that a localized welding of the metallic parts occurs and hence the resistance of the contacts irreversibly drops down. This brilliantly confirms Ashkinass experiments.

In a work published in 2001, Bardou and Boosé theoretically established that the tunnelling probability of a particle through a potential barrier could be strongly enlarged by slightly striking the particle at the time when the centroid of its wave packet is reflecting on the barrier [2]. The one dimensional model calculation they performed stands for a square potential barrier of height $U_0$ and an electron with energy E described as a wave packet. It was shown that a small momentum transfer to the particle $\Delta p_x$ (such that the energy transfer $\hbar^2 \Delta p_x / 2m << E$, m being the mass of the particle) during this time delay has the dramatic effect of projecting its wave functions on the full set of its eigen functions. This full set of wave functions is made of the stationary wave functions of the undisturbed Schrödinger equation. The result of the momentum transfer is therefore such that the weight of the continuum of wave functions corresponding to the particle energies lying in the continuum far above the potential barrier height becomes preponderant in the final wave function. As the transmission probability of the particle through the barrier is obtained by integrating the contributions of all the wave functions, the effect of the momentum transfer is to increase the transmission probability because of the large contribution of the passing wave functions. These authors have shown that the excitation of the particle at the time of its reflection on a potential barrier increases it transmission probability by several orders of magnitude. Precisely, under the above transfer limit, the transmission of the barrier is proportional to the square of the momentum transfer $T(\Delta p_x) \propto (\Delta p_x)^2$. As a consequence the induced tunnelling effect does not depend on the sign of the momentum transfer: the effect is the same whether the particle is pushed toward the barrier or whether it is pulled away from it. This is strongly different from the regular tunnel effect in which the transmission probability varies as $T(E) \propto \exp(-a\sqrt{U_0 - E})$, a being a constant of proportionality that depends on the thickness of the barrier, and hence increases when the particle is pushed toward the barrier (the energy increases) and decreases when the particle is pulled away (the energy decreases).

One important feature of the induced tunnelling effect lies in the fact that it reaches its maximum amplitude at a time that corresponds to the exact reflection on the barrier. A text-book calculation shows that the reflection of a particle on a barrier is not instantaneous it is related to the time it takes for the evanescent waves to establish inside the barrier [16].

$$\tau = \frac{2m}{\hbar k \sqrt{\kappa_0^2 - k^2}}, \quad \kappa_0^2 = \frac{2mU_0}{\hbar^2}, \quad k^2 = \frac{2mE}{\hbar^2} \tag{1}$$

Under the physical conditions, $U_0 = 2$ eV, $E = 0.5$ eV, this time delay is equal to 0.76 fs. It fixes the maximum duration of the momentum transfer to the electron. Outside this time interval any momentum transfer is inefficient and equivalent to an energy change of the impinging electron in the regular tunnelling [2]. Therefore the external excitation of a particle reflecting on a barrier must contain dynamical structures on this time scale. As soon as the beginning of last century, Abraham and Lemoine experimentally discovered that the duration of an electric spark is less than one nanosecond [17]. Recently Descœudres measured the spectral width of electric sparks to be of the order of 400 nm in the visible range [18]. From this spectral width, a coherence time as short as one femtosecond can be inferred from the standard Fourier time-frequency relationship [19] showing that the condition of shortness in the momentum transfer to the electron can be fulfilled by the electromagnetic field emitted by an electric spark.

At this point it becomes clear that the induced tunnelling effect gives the appropriate frame for an interpretation of the Branly effect. In a granular metallic medium microscopic grains are electrically isolated one from the other by a metal oxide nanometric layer, forming a large population of MIM's. When a voltage is applied to the medium, electrons are accelerated and they do reflect on the potential barriers. At the time of the reflection, these electrons can be kicked forward or backward by the short electromagnetic pulses present in the external electromagnetic field emitted by a spark, a small momentum is transferred to the electrons. The enhanced transmission induced by the momentum transfer produces an increased electrical current, that for some events become large enough to permit a local heating in the metal grains thanks to the Joule effect. Eventually a welding of the grains can occur and when a percolation path has formed the electrical resistance of the medium drops down going from an exponential dependence on the applied voltage to a linear one (Ohm's law).

This theory should be easy to falsify. For instance excitation of a coherer by a smooth, long lasting electromagnetic field should be inefficient as a fast momentum transfer is needed that lasts no longer than the reflection time of the electrons on the oxide barriers in the framework of the induced tunnelling effect. Gold dots deposited on a resistive surface and separated by 1 nm thick vacuum (air) barriers should behave as an instantaneous coherer, because the resistance drop would only be controlled by the induced tunnelling effect and no welding between the dots would possible in that experimental situation.

This physical interpretation can be applied to various other manifestations of resistance changes under an external excitation of a granular medium that were observed in the course of exploring the Branly effect. Auerbach demonstrated in 1898 that a coherer could be made conducting by an acoustic excitation in the audible range of the spectrum [20]. As mentioned in Ref. [2] and more precisely described in Ref. [14] there exist another way to produce induced tunnel effect, which rests on the fact that particles get additional momentum too when they impinge on a moving potential barrier [21]. Therefore acoustical wave, by inducing vibration of the tunnel barriers between the grains of a coherer, could be responsible of an induced tunnelling. More, acoustical induced tunnelling could also be at the very origin of the microphone effect between carbon grains that has only been abandoned during the last decade by the telephone industry [22]. Related to these experiments an increase of resistance was observed when metallic filings were replaced by what is nowadays called semiconductor grains. A typical experiment was performed by Ashkinass using $PbO_2$: when externally excited by a strong electrical spark the resistance of a tube filled with lead dioxide powder increases by several orders of magnitude [11]. Here again tunnelling induced by an external electromagnetic wave could be responsible of a charge (electrons and holes) migration on both sides of tunnels separating the grains. The charge separation by creating a region of space charge would increase the thickness of the tunnels and therefore increase the resistance of the medium.

The new concept of induced tunnelling has a strong unifying power of various effects that take place in situations where quantum tunnels are present. As it has been described above, the effect is sensitive to any surrounding mechanical and electromagnetic noise excitations. The resulting consequence is a noisy intrinsic behaviour of any device that can exhibit the Branly effect. As a matter of fact coherers used in the early stages of radio telegraphy have been difficult to handle because of large variation of their output, carbon granular microphones are known to exhibit the specific "carbon hiss" noise, and large noise levels are observed in MIM structures [11,23] used nowadays in memory devices. Since the first measurement, in 1995 of the electrical resistance, of a single $C_{60}$ molecule [24], the growing number of experiments in the field of molecular electronics has been exponential. A large number of these experiments rely on the use of tunnel effects and should therefore be sensitive to induced tunnelling, care should be taken to strictly isolate the experimental set ups from external excitations, electromagnetic or acoustic. Last but not least, it is known for a long time that alpha radioactivity [25,26] is due to the tunnelling of a hydrogen nucleus out of large and unstable nuclei. This consideration opens the opportunity of exploring the possibility for gamma rays with energy less than the barrier height to induce tunnelling of hydrogen particles, hence decreasing the radioactive lifetime of the emitting nuclei. This could also be explored in case of beta emission.

**In memoriam.** This work is dedicated to François Bardou†.

## REFERENCES

[1] É. Branly "Variations de conductibilité sous diverses influences électriques", C. r. hebd. Séanc. Acad. Sci., Paris, **11**, 785-787, (1890).
[2] D. Boosé and F. Bardou, "A quantum evaporation effect," Europhys. Lett., **53**, 1-7 (2001).
[3] K.E. Guthe and A. Trowbridge, "On the theory of the coherer", Phys. Rev., **11**, 22-39 (1900).
[4] T. Tommasina, "Sur un curieux phénomène d'adhérence des limailles métalliques sous l'action du champ électrique" C. r. hebd. Séanc. Acad. Sci., Paris **127** 1014-1016 (1898).
[5] T. Tommasina, "Sur un cohéreur très sensible, obtenu par le simple contact de deux charbons …" C. r. hebd. Séanc. Acad. Sci., Paris, **128**, 1092-1095, (1899).
[6] O. J. Lodge, On the sudden acquisition of conducting-power by a series of discrete metallic particles, Phil. Mag., **37**, 94-95 (1894).
[7] O.J. Lodge, "The history of the coherer principle", The Electrician, **40**, 86-91 (1897).
[8] É. Branly, "Résistance électrique au contact de deux disques d'un meme metal", C. r. hebd. Séanc. Acad. Sci., Paris **127**, 219-221 (1898).
[9] O. Rochefort, Remarque sur le fonctionnement des cohéreurs et des auto-décohéreurs, C. r. hebd. Séanc. Acad. Sci., Paris, **134**, 830-831 (1902).
[10] Falcon E., B. Castaing, and M. Creyssels, "Nonlinear electrical conductivity in a 1D granular medium," Eur. Phys. J. B, **38**, 475-483, 2004.
[11] Ashkinass, E. "Theoretishes und Experimentelles über den Cohärer." Annalen der Physik: (Leipzig) **302**, 284-307 (1898).
[12] Vandembroucq D., A. C. Boccara, and S. Roux, "Breakdown patterns in Branly's coheror," J. Phys. III, France, **7**, 303-310, (1997).
[13] Dorbolo S., Ausloos M., and N. Vandewalle, "Reexamination of the Branly effect," Phys. Rev. E,

**67**, 040302(1-4) (2003).

[14] Ch. Hirlimann, B. Thomas, D. Boosé, Induced optical tunneling, Europhys. Lett. **69**, 48-54 (2005).

[15] Branly, É. "Radioconducteurs à billes métalliques." C. r. hebd. Séanc. Acad. Sci., Paris **128**, 1089-1092 (1899).

[16] C. Cohen-Tannoudji, B. Diu, and F. Laloë, in Mécanique quantique Vol. 1, Hermann, paris, 1977.

[17] H. Abraham and J. Lemoine, Ann. Chim. Phys. (Paris), **20**, 264-287 (1900).

[18] A. Decœudres, Thesis N° 3542 de l'École Polytechnique Fédérale de Lausanne, "Characterization of electrical discharge machining plasma", Lausanne EPFL (2006) p. 67-75.

[19] C. Hirlimann in "Femtosecond Laser pulses, principles and Experiments", 2nd edition, C. Rullière ed. (Springer-Verlag, 2005) p. 31-32

[20] Auerbach, F. "Ueber Widerstandsverminderung duch electrische und durch akustische Schwingungen." Annalen der Physik: (Leipzig) **300**, 611-617 (1898).

[21] Landau L. and Lifchitz E., Mécanique Quantique, Théorie non relativiste (Mir, Moscou) 1967, pp. 174, 175.

[22] The carbon microphone was invented in 1878 by David Edward Hughes.

[23] V. Da Costa, M. Romeo, F. Bardou, Statistical properties of currents flowing through tunnel junctions, J. Magn. Magn. Mat. **258-259**, 90-95 (2003).

[24] C. Joachim, J. K. Gimzewski, R. R. Schlittler, C. Chavy, Electronic Transparence of a Single $C_{60}$ Molecule, Phys. Rev. Lett., **74**, 2102 - 2105 (1995).

[25] G. Gamow, Zur quantentheorie des atomkernes, Z. Phys. A, **51**, 204-212 (1928).

[26] R. W. Gurney, E. U. Condon, Quantum mechanics and radioactive desintegration, Phys. Rev. **33**, 127-140 (1929).